\documentclass[aps,prl,twocolumn,showpacs,amsmath,preprintnumbers,amssymb,groupedaddress,superscriptaddress]{revtex4}
\usepackage{graphicx}
\usepackage{color}

\newcommand{\Jpsi}{$J/\psi$ }
\newcommand{\pT}{$p_T$ }
\newcommand{\xT}{$x_T$ }
\newcommand{\sNN}{$\sqrt{s_{_\mathrm{NN}}}$ }
\newcommand{\s}{$\sqrt{s}$ }
\newcommand{\pp}{$p$+$p$ }
\newcommand{\cucu}{Cu+Cu }
\newcommand{\raa}{$R_{AA}$ }

\begin{document}


\title{\Jpsi production at high transverse momenta in \pp and \cucu collisions at \sNN = 200 GeV}

%
%

\affiliation{Argonne National Laboratory, Argonne, Illinois 60439,
USA} \affiliation{University of Birmingham, Birmingham, United
Kingdom} \affiliation{Brookhaven National Laboratory, Upton, New
York 11973, USA} \affiliation{University of California, Berkeley,
California 94720, USA} \affiliation{University of California,
Davis, California 95616, USA} \affiliation{University of
California, Los Angeles, California 90095, USA}
\affiliation{Universidade Estadual de Campinas, Sao Paulo, Brazil}
\affiliation{University of Illinois at Chicago, Chicago, Illinois
60607, USA} \affiliation{Creighton University, Omaha, Nebraska
68178, USA} \affiliation{Nuclear Physics Institute AS CR, 250 68
\v{R}e\v{z}/Prague, Czech Republic} \affiliation{Laboratory for
High Energy (JINR), Dubna, Russia} \affiliation{Particle Physics
Laboratory (JINR), Dubna, Russia} \affiliation{Institute of
Physics, Bhubaneswar 751005, India} \affiliation{Indian Institute
of Technology, Mumbai, India} \affiliation{Indiana University,
Bloomington, Indiana 47408, USA} \affiliation{Institut de
Recherches Subatomiques, Strasbourg, France}
\affiliation{University of Jammu, Jammu 180001, India}
\affiliation{Kent State University, Kent, Ohio 44242, USA}
\affiliation{University of Kentucky, Lexington, Kentucky,
40506-0055, USA} \affiliation{Institute of Modern Physics,
Lanzhou, China} \affiliation{Lawrence Berkeley National
Laboratory, Berkeley, California 94720, USA}
\affiliation{Massachusetts Institute of Technology, Cambridge, MA
02139-4307, USA} \affiliation{Max-Planck-Institut f\"ur Physik,
Munich, Germany} \affiliation{Michigan State University, East
Lansing, Michigan 48824, USA} \affiliation{Moscow Engineering
Physics Institute, Moscow Russia} \affiliation{City College of New
York, New York City, New York 10031, USA} \affiliation{NIKHEF and
Utrecht University, Amsterdam, The Netherlands} \affiliation{Ohio
State University, Columbus, Ohio 43210, USA} \affiliation{Old
Dominion University, Norfolk, VA, 23529, USA} \affiliation{Panjab
University, Chandigarh 160014, India} \affiliation{Pennsylvania
State University, University Park, Pennsylvania 16802, USA}
\affiliation{Institute of High Energy Physics, Protvino, Russia}
\affiliation{Purdue University, West Lafayette, Indiana 47907,
USA} \affiliation{Pusan National University, Pusan, Republic of
Korea} \affiliation{University of Rajasthan, Jaipur 302004, India}
\affiliation{Rice University, Houston, Texas 77251, USA}
\affiliation{Universidade de Sao Paulo, Sao Paulo, Brazil}
\affiliation{University of Science \& Technology of China, Hefei
230026, China} \affiliation{Shandong University, Jinan, Shandong
250100, China} \affiliation{Shanghai Institute of Applied Physics,
Shanghai 201800, China} \affiliation{SUBATECH, Nantes, France}
\affiliation{Texas A\&M University, College Station, Texas 77843,
USA} \affiliation{University of Texas, Austin, Texas 78712, USA}
\affiliation{Tsinghua University, Beijing 100084, China}
\affiliation{United States Naval Academy, Annapolis, MD 21402,
USA} \affiliation{Valparaiso University, Valparaiso, Indiana
46383, USA} \affiliation{Variable Energy Cyclotron Centre, Kolkata
700064, India} \affiliation{Warsaw University of Technology,
Warsaw, Poland} \affiliation{University of Washington, Seattle,
Washington 98195, USA} \affiliation{Wayne State University,
Detroit, Michigan 48201, USA} \affiliation{Institute of Particle
Physics, CCNU (HZNU), Wuhan 430079, China} \affiliation{Yale
University, New Haven, Connecticut 06520, USA}
\affiliation{University of Zagreb, Zagreb, HR-10002, Croatia}

\author{B.~I.~Abelev}\affiliation{University of Illinois at Chicago, Chicago, Illinois 60607, USA}
\author{M.~M.~Aggarwal}\affiliation{Panjab University, Chandigarh 160014, India}
\author{Z.~Ahammed}\affiliation{Variable Energy Cyclotron Centre, Kolkata 700064, India}
\author{B.~D.~Anderson}\affiliation{Kent State University, Kent, Ohio 44242, USA}
\author{D.~Arkhipkin}\affiliation{Particle Physics Laboratory (JINR), Dubna, Russia}
\author{G.~S.~Averichev}\affiliation{Laboratory for High Energy (JINR), Dubna, Russia}
\author{J.~Balewski}\affiliation{Massachusetts Institute of Technology, Cambridge, MA 02139-4307, USA}
\author{O.~Barannikova}\affiliation{University of Illinois at Chicago, Chicago, Illinois 60607, USA}
\author{L.~S.~Barnby}\affiliation{University of Birmingham, Birmingham, United Kingdom}
\author{J.~Baudot}\affiliation{Institut de Recherches Subatomiques, Strasbourg, France}
% ...

\collaboration{STAR Collaboration}\noaffiliation

\date{\today}
\begin{abstract}
The STAR collaboration at RHIC presents measurements of
\Jpsi$\rightarrow{e^+e^-}$ at mid-rapidity and high transverse
momentum ($p_T>5$ GeV/$c$) in \pp and central \cucu collisions at
\sNN = 200 GeV. The inclusive \Jpsi\ production cross section for
\cucu collisions  is found to be consistent at high $p_T$ with the
binary collision-scaled cross section for \pp collisions, in
contrast to previous measurements at lower $p_T$, where a
suppression of \Jpsi production is observed relative to the
expectation from binary scaling. Azimuthal correlations of
$J/\psi$ with charged hadrons in \pp collisions provide an
estimate of the contribution of $B$-meson decays to \Jpsi
production of $13\% \pm 5\%$.
\end{abstract}



\maketitle


Suppression of the $c\bar{c}$ bound state \Jpsi meson production
in relativistic heavy-ion collisions arising from \Jpsi
dissociation due to screening of the $c\bar{c}$ binding potential
in the deconfined medium has been proposed as a signature of
Quark-Gluon Plasma (QGP) formation~\cite{colorscreen}.
Measurements at \sNN $=17.3$ GeV at the CERN-SPS observed a strong
suppression of \Jpsi production in heavy-ion
collisions~\cite{Abreu:2000xe}, although the magnitude of the
suppression decreases with increasing \Jpsi $p_T$. This systematic
dependence may be explained by initial state scattering (Cronin
effect \cite{two_component_approach}), as well as the combined
effects of finite \Jpsi\ formation time and the finite space-time
extent of the hot, dense volume where the dissociation can occur
\cite{Karsch:1988ri}.

At higher beam energy (\sNN $= 200$ GeV), the PHENIX collaboration
at RHIC has measured $J/\psi$ suppression for $p_T < 5$ GeV/$c$ in
central (small impact parameter) Au+Au and Cu+Cu
collisions~\cite{Adare:2006ns} that is similar in magnitude to
that observed at the CERN-SPS. This similarity is surprising in
light of the expectation that the energy density is significantly
higher at larger collision energy. It may be due to the
counterbalancing of larger dissociation with recombination of
unassociated $c$ and $\bar{c}$ in the medium, which are more
abundant at higher energy
\cite{BraunMunzinger:2000px,Grandchamp:2001pf,Gorenstein:2001xp,Thews:2000rj,Frawley:2008kk}.

Measurements of open heavy-flavor production may also shed light
on $J/\psi$ suppression mechanisms. Non-photonic electrons from
the semi-leptonic decay of heavy flavor mesons are found to be
strongly suppressed in heavy-ion relative to p+p collisions at
RHIC \cite{STAR_NPE,PHENIX_NPE}, an effect that has been
attributed to partonic energy loss in dense matter
\cite{Dokshitzer:2001zm}. This process may also contribute to
high-$p_T$ \Jpsi suppression, if \Jpsi formation proceeds through
a channel carrying color.

The medium generated in RHIC heavy-ion collisions is thought to be
strongly coupled \cite{STAR_whitePaper}, making accurate QCD
calculations of quarkonium propagation difficult. The AdS/CFT
duality for QCD-like theories may provide insight into heavy
fermion pair propagation in a strongly coupled liquid. One such
calculation predicts that the dissociation temperature decreases
with increasing \Jpsi \pT (or velocity)~\cite{adscft}. The
temperature achieved at RHIC ($\sim 1.5$
T${}_c$)~\cite{STAR_whitePaper} is below this dissociation
temperature at low \Jpsi $p_T$, and above it at $p_T\gtrsim5$
GeV/$c$. Consequently, \Jpsi production is predicted to be more
suppressed at high $p_T$, in contrast to the standard suppression
mechanism. This prediction can be tested with measurements of
\Jpsi over a broad kinematic range, in both \pp and nuclear
collisions.

The interpretation of \Jpsi suppression observed at the SPS and by
the PHENIX collaboration requires understanding of the quarkonium
production mechanism in hadronic collisions, which include direct
production via gluon fusion and color-octet (CO) and color-singlet
(CS) transitions, as described by Non-Relativistic Quantum
ChromoDynamics (NRQCD) \cite{Bodwin:1994jh}; parton fragmentation;
and feeddown from higher charmonium states ($\chi_c$, $\psi(2S)$)
and $B$ meson decays. No model at present fully explains the \Jpsi
systematics observed in elementary collisions
\cite{QWG_YellowReport}. \Jpsi measurements at high-$p_T$ both in
\pp and nuclear collisions may provide additional insights into
the basic processes underlying quarkonium production.


This letter reports new measurements by the STAR collaboration at
RHIC of \Jpsi production at high transverse momentum in \pp and
\cucu collisions at \sNN = 200 GeV \cite{Ackermann:2002ad}. The
inclusive cross section and semi-inclusive $J/\psi$-hadron
correlations are presented.



The \cucu data are from the RHIC 2005 run, while the \pp data are
from 2005 and 2006. The online trigger, utilizing the STAR Barrel
Electromagnetic Calorimeter (BEMC) \cite{STAR_BEMC} as well as
other trigger detectors, required one BEMC tower with an energy
deposition above a given threshold in coincidence with a minimum
bias (MB) collision trigger~\cite{STAR_embedding}. The online
trigger threshold, MB trigger condition, and sampled integrated
luminosity for each dataset are listed in Tab.~\ref{table:cuts}.
In \cucu data, the most central 0-20\% and 0-60\% of the total
hadronic cross section were selected as
in~\cite{STAR_embedding,bedangaPhi}.

In this analysis, \Jpsi$\rightarrow{e^+e^-}$ (Branching Ratio
(B)=5.9\%) was reconstructed using the STAR Time Projection
Chamber (TPC) \cite{STAR_TPC} and BEMC, with acceptance $|\eta|<1$
and full azimuthal coverage. Hadron rejection was achieved through
the combination of BEMC shower energy, shower shape measured in
the embedded Shower-Maximum Detector (SMD), and ionization loss
($dE/dx$) in the TPC~\cite{STAR_NPE,TPCReCalib}. Electron purity
is $>70\%$ with high efficiency.  At moderate $p_T$, the TPC alone
can measure electrons with efficiency $>90\%$ and sufficient
hadron rejection ($\sim10^3$)~\cite{STAR_NPE,starTOFelectron}.


\begin{table}
\caption{Trigger conditions, off-line cuts and \Jpsi signal
statistics. $E_T$ is the BEMC trigger threshold. $p_{T1}$ and
$p_{T2}$ are the lower bounds for the two electron candidates. BBC
(ZDC) means the coincidence of Beam Beam Counters (Zero Degree
Calorimeters). S/B is the ratio of signal to background.
\label{table:cuts}}
\begin{tabular}{cccc}
\hline \hline
& \pp (2005) & \pp (2006) & \cucu \\
\hline
MB trigger & BBC & BBC & ZDC \\
$E_{T}$ (GeV)& $>$ 3.5  &  $>$ 5.4  & $>$
3.75 \\
Sampled int. lumi   & 2.8 $pb^{-1}$ & 11.3 $pb^{-1}$ & 860
$\mu b^{-1}$ \\
\hline
$p_{T1}$ (GeV/$c$)  & $>$ 2.5 & $>$ 4.0 & $>$ 3.5\\
$p_{T2}$ (GeV/$c$)  & $>$ 1.2 & $>$ 1.2 & $>$ 1.5 \\
\Jpsi \pT (GeV/c) & 5-8  & 5-14 & 5-8\\
\hline
\Jpsi counts        & 32 $\pm$ 6 & 51 $\pm$ 10 & 23 $\pm$ 8\\
S/B                 & 9:1 & 2:1 & 1:4 \\

\hline \hline
\end{tabular}
\end{table}



\begin{figure}[tbp]
\centering \includegraphics[width=0.98\columnwidth]{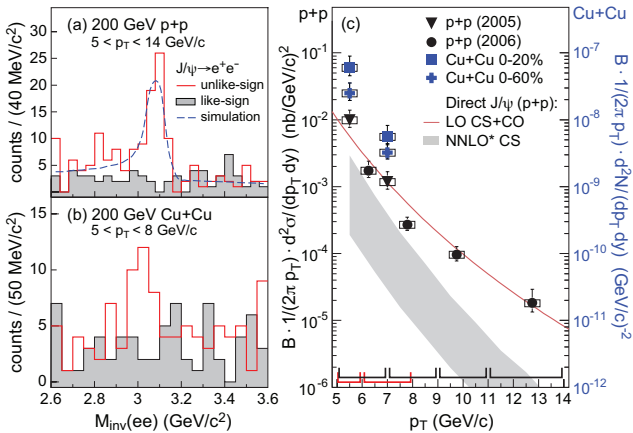}
\caption{(Color online.) Left: invariant dielectron mass
distribution in (a) \pp and (b) \cucu collisions, for opposite
sign (solid red) and same sign pairs (grey band) from data, and
simulated \Jpsi peak for \pp (dashed).  Right: \Jpsi \pT
distributions in \pp and \cucu collisions at \sNN = 200 GeV.
Horizontal brackets show bin limits. Also shown are perturbative
calculations for LO CS+CO (solid line) and NNLO* CS (band) direct
yields, without feeddown contributions.} \label{invm}
\end{figure}

Figure \ref{invm} shows di-electron invariant mass distributions
for (a) \pp and (b) \cucu collisions at \sNN = 200 GeV. The
like-sign distribution measures random pair background from Dalitz
decays and photon conversions. The \Jpsi mass window is
$2.7<M_{inv}^{ee}<3.2$ GeV/$c^2$. Other correlated $e^+e^-$
background is estimated to be
$<10\%$~\cite{Adare:2006kf,JpsiSpectra_CDF,JpsiSpectra_CDFII}.
Table~\ref{table:cuts} lists the offline cuts and \Jpsi signal
statistics. Different thresholds were used for the two electron
candidates, corresponding to different online trigger thresholds.

The \Jpsi detection efficiency was calculated by two complementary
methods. The first method was to determine the electron trigger
efficiency by comparing triggered electron yield to the measured
inclusive electron spectrum \cite{STAR_NPE}. The non-triggered
electron efficiency depends only on the TPC tracking efficiency,
which was determined by embedding simulated electron tracks into
real events~\cite{STAR_embedding}, and $dE/dx$ efficiencies,
determined from the distributions in real data~\cite{TPCReCalib}.
The second method was to simulate $J/\psi$ events in
PYTHIA~\cite{Sjostrand:2006za}, embed them into real events, and
reconstruct the hybrid event to determine the \Jpsi trigger and
detection efficiencies. The difference in estimated efficiency
between the two methods is $<10\%$ for all datasets and is
included into the systematic uncertainties of the inclusive
spectra. This systematic uncertainty is correlated in \pp and
Cu+Cu. A log-likelihood method is used to correct the \Jpsi
efficiency and calculate the yields~\cite{zeboThesis}.

Figure~\ref{invm} (c) shows the measured
$J/\psi\rightarrow{e^+e^-}$ \pT spectra. The systematic
uncertainties are dominated by kinematic cuts, trigger efficiency
(9\%) and reconstruction efficiency (8\%), and are similar and
correlated in \pp and Cu+Cu. The normalization uncertainty for the
inclusive non-singly diffractive \pp cross section is
14\%~\cite{ppUncertainty}. Theoretical calculations shown in the
figure are NRQCD from CO and CS transitions for direct $J/\psi$'s
in $p+p$ collisions \cite{Nayak:2003jp} (solid line) and
NNLO$^\star$ CS result \cite{Artoisenet:2008fc} (gray band).
Neither calculation includes feeddown contributions. The band for
NNLO$^\star$ gives the uncertainty due to scale parameters and the
charm quark mass. The CS+CO calculation describes the data well
and leaves little room for feeddown from $\psi'$, $\chi_c$ and
$B$, estimated to be a factor of $\sim1.5$. NNLO$^\star$ CS
predicts a steeper \pT dependence.


\begin{figure}[tbp]
\centering
\includegraphics[width=0.95\columnwidth]{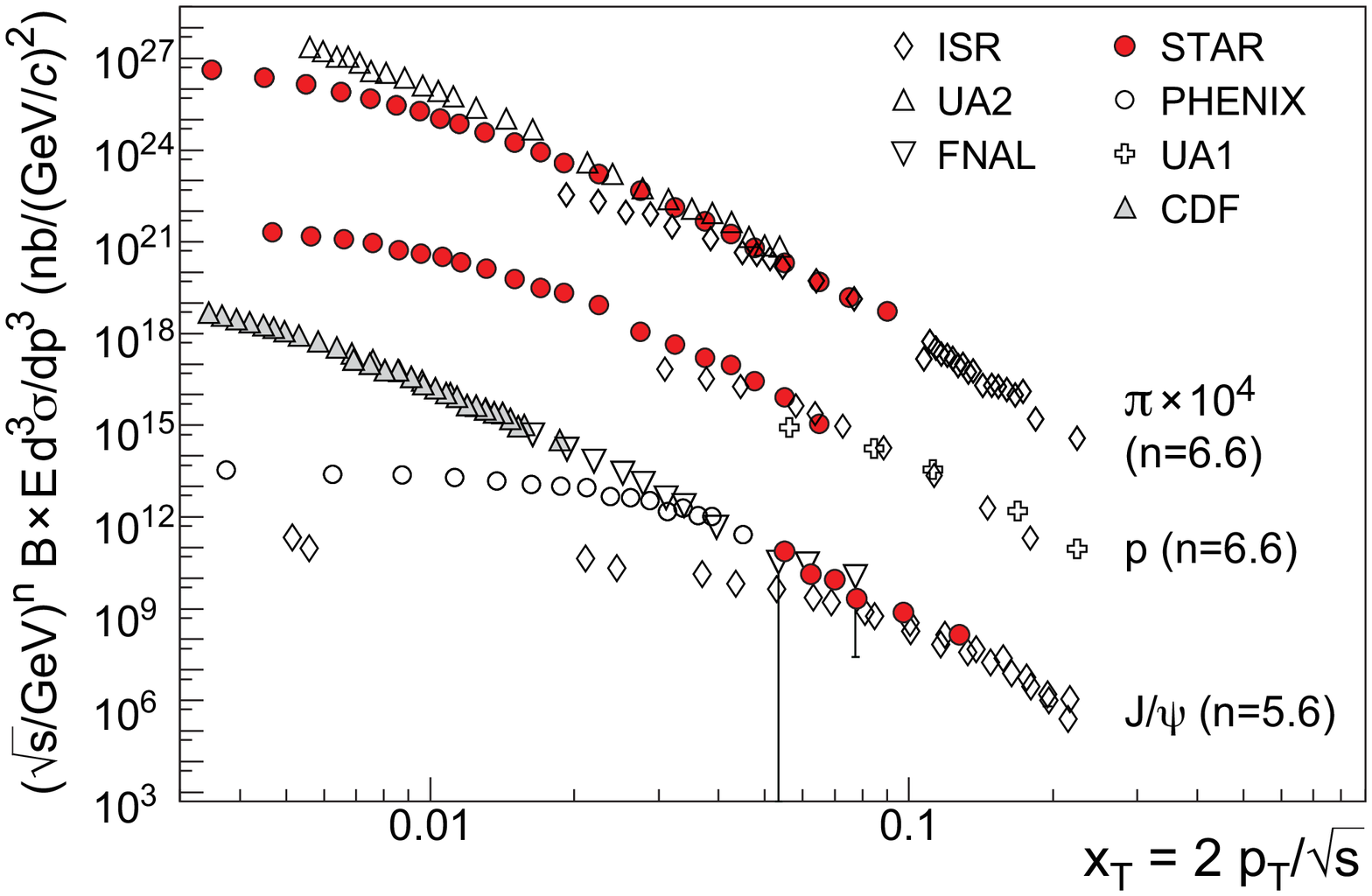}
\caption{ \xT distributions of pions and protons
\cite{pion_UA2,scalingpi,Lijuan,pion_proton_ISR,proton_FNAL}
and $J/\psi$ (CDF \cite{JpsiSpectra_CDFII, JpsiSpectra_CDF}, UA1
\cite{UA1Simulation}, PHENIX \cite{Adare:2006kf}, and ISR
\cite{JpsiSpectra_ISR}).} \label{scaling}
\end{figure}

Proton and pion inclusive production cross sections in high energy
\pp collisions have been found to follow \xT scaling
\cite{xT_scaling_history1,xT_scaling_history2,xT_scaling_history3}:
$E\frac{d^3\sigma}{dp^3}=g(x_T)/s^{n/2}$, where
$x_T=2p_T/\sqrt{s}$. In the parton model, $n$ reflects the number
of constituents taking an active role in hadron production.
Figure~\ref{scaling} shows the \xT distributions of this data and
previous $J/\psi$, pion and proton data, from \pp collisions. The
\Jpsi data \cite{JpsiSpectra_CDFII,JpsiSpectra_CDF, UA1Simulation,
Adare:2006kf,JpsiSpectra_ISR} cover the range \s=30 GeV to \s=1.96
TeV. The $J/\psi$ exhibits \xT scaling ($n=5.6\pm0.2$) at high
$p_T$, similar to the trend for pions and protons ($n=6.6\pm0.1$)
\cite{scalingpi,Lijuan}. While low \pT \Jpsi production originates
in a hard process due to the mass scale, subsequent soft processes
could cause violation of \xT scaling. At high $p_T$, the power
parameter $n=5.6\pm0.2$ is closer to the predictions from CO and
Color-Evaporation production
($n\simeq6$)~\cite{Nayak:2003jp,Bedjidian:2004gd} and much smaller
than that from next-to-next-to leading order (NNLO*) CS production
($n\simeq8$)~\cite{Artoisenet:2008fc}. This is also evident from
Fig.~\ref{invm} (c).

\begin{figure}[tbp]
\centering
\includegraphics[width=0.95\columnwidth]{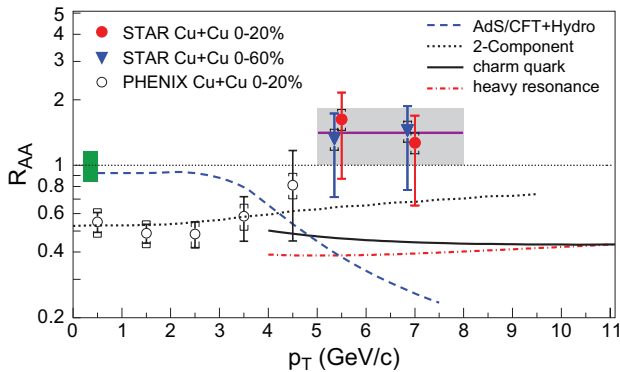}
\caption{\label{raa} (Color online). \Jpsi \raa vs. $p_T$. STAR
data points have statistical (bars) and systematic (caps)
uncertainties. The box about unity on the left shows $R_{AA}$
normalization uncertainty, which is the quadrature sum of p+p
normalization and binary collision scaling uncertainties.  The
solid line and band show the average and uncertainty of the two
0-20\% data points. The curves are model calculations described in
the text. The uncertainty band of 10\% for the dotted curve is not
shown. }
\end{figure}

The nuclear modification factor $R_{AA}(p_T$) \cite{Adler:2002xw},
defined as the ratio of the inclusive hadron yield in nuclear
collisions to that in \pp collisions scaled by the underlying
number of binary nucleon-nucleon collisions, measures
medium-induced effects on inclusive particle production. In the
absence of such effects, \raa is unity for hard processes.

Figure \ref{raa} shows \raa for \Jpsi vs $p_T$, in 0-20\% \cucu
collisions from PHENIX \cite{PHENIX_CuCu} and STAR, and 0-60\%
\cucu from STAR. \cucu and \pp data with $p_T>5$ GeV/$c$ are from
STAR. The \raa systematic uncertainty takes into account the
correlated efficiencies of the \cucu and \pp datasets. \raa for
\Jpsi is seen to increase with increasing $p_T$. The average of
the two STAR 0-20\% data points at high-\pT is
$R_{AA}=1.4\pm0.4~(stat.)\pm0.2~(syst.)$. Utilizing the STAR \cucu
and \pp data reported here and PHENIX \cucu data at high-\pT
\cite{PHENIX_CuCu} gives $R_{AA} =
1.1\pm0.3~(stat.)\pm0.2~(syst.)$ for $p_T>5$ GeV/$c$.  Both
results are consistent with unity and differ by two standard
deviations from a PHENIX measurement at lower \pT
($R_{AA}=0.52\pm0.05$ \cite{PHENIX_CuCu}). A notable conclusion
from these data is that \Jpsi is the only hadron measured in RHIC
heavy-ion collisions that does not exhibit significant high \pT
suppression. However, for the \Jpsi population reported here, the
initial scattered partons have average momentum fraction $\sim
0.1$ (see also Fig. 2), where initial state effects such as
anti-shadowing may lead to increasing $R_{AA}$ with increasing
$p_T$.

The dashed curve in Fig.~\ref{raa} shows the prediction of an
AdS/CFT-based calculation, in which the \Jpsi is embedded in a
hydrodynamic model~\cite{hydro+hotWind} and the \Jpsi dissociation
temperature decreases with increasing velocity according
to~\cite{adscft}. Its \pT dependence is at variance with that of
the data.  The dotted line shows the prediction of a two-component
model including color screening, hadronic phase dissociation,
statistical $c\bar{c}$ coalescence at the hadronization
transition, \Jpsi formation time effects, and $B$-meson feeddown
\cite{two_component_approach}. This calculation describes the
overall trend of the data.

The other calculations in Fig.~\ref{raa} provide a comparison to
open charm $R_{AA}$. The solid line is based on the WHDG model for
charm quark energy loss, with assumed medium gluon density
$dN_g/dy = 254$ for 0-20\% Cu+Cu \cite{Wicks:2005gt}. The
dash-dotted line shows a GLV model calculation for D-meson energy
loss, with $dN_g/dy = 275$ \cite{Adil:2006ra}. Both models, which
correctly describe heavy-flavor suppression in Au+Au collisions,
predict charm meson suppression of a factor $\sim2$ at $p_T > 5$
GeV/$c$. This is in contrast to the \Jpsi $R_{AA}$. This
comparison suggests that high-$p_T$ \Jpsi production does not
proceed dominantly via a channel carrying color. However, other
effects~\cite{two_component_approach,xmXu} may compensate for the
predicted loss in this \pT range.


\begin{figure}[tbp]
\centering
\includegraphics[width=0.95\columnwidth]{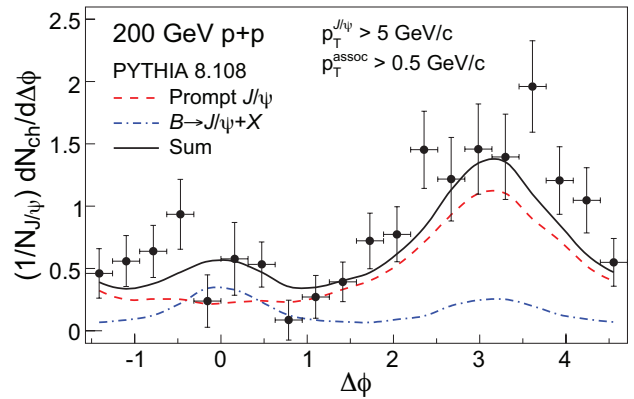}
\caption{\label{corr} (Color online). $J/\psi$-hadron azimuthal
correlations.  Lines show PYTHIA calculation of prompt (dashed)
and $B$-meson (dot-dashed) feeddown contributions, and their sum
(solid).}
\end{figure}

Figure~\ref{corr} shows the azimuthal correlation between high-\pT
\Jpsi ($p_T>5$ GeV/$c$) and charged hadrons with $p_T>0.5$ GeV/$c$
in 200 GeV p+p collisions. The \Jpsi mass window is narrowed to
2.9-3.2 GeV/$c^2$ to increase the S/B ratio. There is no
significant correlated yield in the near-side ($\Delta \phi \sim
0$), in contrast to dihadron correlation measurements
\cite{STAR_diHadron}. The lines show the result of a PYTHIA
calculation \cite{Sjostrand:2006za}, which exhibits a near-side
correlation due dominantly to $B \rightarrow J/\psi + X$. A
$\chi^2$ fit to the data of the summed distribution (directly
produced \Jpsi, feeddown from $\chi_c$, $\psi(2S)$ and $B$-meson)
gives a contribution from $B$-meson feeddown to inclusive \Jpsi
production of $13\% \pm 5\%$ at $p_T>5$ GeV/$c$.


In summary, we report new measurements of \Jpsi production in
\s=200 GeV \pp and \cucu collisions at high \pT ($p_T>5$ GeV/$c$)
at RHIC. The \Jpsi inclusive cross section was found to obey \xT
scaling for \pT$\gtrsim5$ GeV/c, in contrast to lower \pT \Jpsi
production. The \Jpsi nuclear modification factor \raa in \cucu
increases from low to high $p_T$ and is consistent with no \Jpsi
suppression for \pT$>$5 GeV/c, in contrast to the prediction from
a theoretical model of quarkonium dissociation in a strongly
coupled liquid using an AdS/CFT approach. The two-component model
with finite \Jpsi formation time describes the increasing trend of
the \Jpsi $R_{AA}$. Based on the measurement of azimuthal
correlations and the comparison to model calculations, we estimate
the fraction of \Jpsi from $B$-meson decay to be $13\pm5\%$ at
$p_T>5$ GeV/$c$.


The authors thank G.C. Nayak, J.P. Lansberg, W.A. Horowitz and I.
Vitev for providing calculations and discussion. We thank the RHIC
Operations Group and RCF at BNL, and the NERSC Center at LBNL and
the resources provided by the Open Science Grid consortium for
their support. This work was supported in part by the Offices of
NP and HEP within the U.S.~DOE Office of Science, the U.S.~NSF,
the Sloan Foundation, the DFG Excellence Cluster EXC153 of
Germany, CNRS/IN2P3, RA, RPL, and EMN of France, STFC and EPSRC of
the United Kingdom, FAPESP of Brazil, the Russian Ministry of Sci.
and Tech., the NNSFC, CAS, MoST, and MoE of China, IRP and GA of
the Czech Republic, FOM of the Netherlands, DAE, DST, and CSIR of
the Government of India, the Polish State Committee for Scientific
Research,  and the Korea Sci. \& Eng. Foundation.
\bibliography{highPtJpsi}

\begin{thebibliography}{50}
\expandafter\ifx\csname natexlab\endcsname\relax\def\natexlab#1{#1}\fi
\expandafter\ifx\csname bibnamefont\endcsname\relax
  \def\bibnamefont#1{#1}\fi
\expandafter\ifx\csname bibfnamefont\endcsname\relax
  \def\bibfnamefont#1{#1}\fi
\expandafter\ifx\csname citenamefont\endcsname\relax
  \def\citenamefont#1{#1}\fi
\expandafter\ifx\csname url\endcsname\relax
  \def\url#1{\texttt{#1}}\fi
\expandafter\ifx\csname urlprefix\endcsname\relax\def\urlprefix{URL }\fi
\providecommand{\bibinfo}[2]{#2}
\providecommand{\eprint}[2][]{\url{#2}}

\bibitem[{\citenamefont{Matsui and Satz}(1986)}]{colorscreen}
\bibinfo{author}{\bibfnamefont{T.}~\bibnamefont{Matsui}} \bibnamefont{and}
  \bibinfo{author}{\bibfnamefont{H.}~\bibnamefont{Satz}},
  \bibinfo{journal}{Phys. Lett.} \textbf{\bibinfo{volume}{B178}},
  \bibinfo{pages}{416} (\bibinfo{year}{1986}).

\bibitem[{\citenamefont{Abreu et~al.}(2001)}]{Abreu:2000xe}
\bibinfo{author}{\bibfnamefont{M.~C.} \bibnamefont{Abreu}}
  \bibnamefont{et~al.}, \bibinfo{journal}{Phys. Lett.}
  \textbf{\bibinfo{volume}{B499}}, \bibinfo{pages}{85} (\bibinfo{year}{2001}).

\bibitem[{\citenamefont{Zhao and Rapp}(2007)}]{two_component_approach}
\bibinfo{author}{\bibfnamefont{X.}~\bibnamefont{Zhao}} \bibnamefont{and}
  \bibinfo{author}{\bibfnamefont{R.}~\bibnamefont{Rapp}}
  (\bibinfo{year}{2007}), \eprint{arXiv:0712.2407}.

\bibitem[{\citenamefont{Karsch and Petronzio}(1988)}]{Karsch:1988ri}
\bibinfo{author}{\bibfnamefont{F.}~\bibnamefont{Karsch}} \bibnamefont{and}
  \bibinfo{author}{\bibfnamefont{R.}~\bibnamefont{Petronzio}},
  \bibinfo{journal}{Phys. Lett.} \textbf{\bibinfo{volume}{B212}},
  \bibinfo{pages}{255} (\bibinfo{year}{1988}).

\bibitem[{\citenamefont{Adare et~al.}(2007{\natexlab{a}})}]{Adare:2006ns}
\bibinfo{author}{\bibfnamefont{A.}~\bibnamefont{Adare}} \bibnamefont{et~al.},
  \bibinfo{journal}{Phys. Rev. Lett.} \textbf{\bibinfo{volume}{98}},
  \bibinfo{pages}{232301} (\bibinfo{year}{2007}{\natexlab{a}}).

\bibitem[{\citenamefont{Braun-Munzinger and
  Stachel}(2000)}]{BraunMunzinger:2000px}
\bibinfo{author}{\bibfnamefont{P.}~\bibnamefont{Braun-Munzinger}}
  \bibnamefont{and} \bibinfo{author}{\bibfnamefont{J.}~\bibnamefont{Stachel}},
  \bibinfo{journal}{Phys. Lett.} \textbf{\bibinfo{volume}{B490}},
  \bibinfo{pages}{196} (\bibinfo{year}{2000}).

\bibitem[{\citenamefont{Grandchamp and Rapp}(2001)}]{Grandchamp:2001pf}
\bibinfo{author}{\bibfnamefont{L.}~\bibnamefont{Grandchamp}} \bibnamefont{and}
  \bibinfo{author}{\bibfnamefont{R.}~\bibnamefont{Rapp}},
  \bibinfo{journal}{Phys. Lett.} \textbf{\bibinfo{volume}{B523}},
  \bibinfo{pages}{60} (\bibinfo{year}{2001}).

\bibitem[{\citenamefont{Gorenstein et~al.}(2002)}]{Gorenstein:2001xp}
\bibinfo{author}{\bibfnamefont{M.~I.} \bibnamefont{Gorenstein}}
  \bibnamefont{et~al.}, \bibinfo{journal}{Phys. Lett.}
  \textbf{\bibinfo{volume}{B524}}, \bibinfo{pages}{265} (\bibinfo{year}{2002}).

\bibitem[{\citenamefont{Thews et~al.}(2001)\citenamefont{Thews, Schroedter, and
  Rafelski}}]{Thews:2000rj}
\bibinfo{author}{\bibfnamefont{R.~L.} \bibnamefont{Thews}},
  \bibinfo{author}{\bibfnamefont{M.}~\bibnamefont{Schroedter}},
  \bibnamefont{and} \bibinfo{author}{\bibfnamefont{J.}~\bibnamefont{Rafelski}},
  \bibinfo{journal}{Phys. Rev.} \textbf{\bibinfo{volume}{C63}},
  \bibinfo{pages}{054905} (\bibinfo{year}{2001}).

\bibitem[{\citenamefont{Frawley et~al.}(2008)\citenamefont{Frawley, Ullrich,
  and Vogt}}]{Frawley:2008kk}
\bibinfo{author}{\bibfnamefont{A.~D.} \bibnamefont{Frawley}},
  \bibinfo{author}{\bibfnamefont{T.}~\bibnamefont{Ullrich}}, \bibnamefont{and}
  \bibinfo{author}{\bibfnamefont{R.}~\bibnamefont{Vogt}},
  \bibinfo{journal}{Phys. Rept.} \textbf{\bibinfo{volume}{462}},
  \bibinfo{pages}{125} (\bibinfo{year}{2008}).

\bibitem[{\citenamefont{Abelev et~al.}(2007)}]{STAR_NPE}
\bibinfo{author}{\bibfnamefont{B.~I.} \bibnamefont{Abelev}}
  \bibnamefont{et~al.}, \bibinfo{journal}{Phys. Rev. Lett.}
  \textbf{\bibinfo{volume}{98}}, \bibinfo{pages}{192301}
  (\bibinfo{year}{2007}).

\bibitem[{\citenamefont{Adare et~al.}(2007{\natexlab{b}})}]{PHENIX_NPE}
\bibinfo{author}{\bibfnamefont{A.}~\bibnamefont{Adare}} \bibnamefont{et~al.},
  \bibinfo{journal}{Phys. Rev. Lett.} \textbf{\bibinfo{volume}{98}},
  \bibinfo{pages}{172301} (\bibinfo{year}{2007}{\natexlab{b}}).

\bibitem[{\citenamefont{Dokshitzer and Kharzeev}(2001)}]{Dokshitzer:2001zm}
\bibinfo{author}{\bibfnamefont{Y.~L.} \bibnamefont{Dokshitzer}}
  \bibnamefont{and} \bibinfo{author}{\bibfnamefont{D.~E.}
  \bibnamefont{Kharzeev}}, \bibinfo{journal}{Phys. Lett.}
  \textbf{\bibinfo{volume}{B519}}, \bibinfo{pages}{199} (\bibinfo{year}{2001}).

\bibitem[{\citenamefont{Adams et~al.}(2005{\natexlab{a}})}]{STAR_whitePaper}
\bibinfo{author}{\bibfnamefont{J.}~\bibnamefont{Adams}} \bibnamefont{et~al.},
  \bibinfo{journal}{Nucl. Phys.} \textbf{\bibinfo{volume}{A757}},
  \bibinfo{pages}{102} (\bibinfo{year}{2005}{\natexlab{a}}).

\bibitem[{\citenamefont{Liu et~al.}(2007)\citenamefont{Liu, Rajagopal, and
  U.A.Wiedemann}}]{adscft}
\bibinfo{author}{\bibfnamefont{H.}~\bibnamefont{Liu}},
  \bibinfo{author}{\bibfnamefont{K.}~\bibnamefont{Rajagopal}},
  \bibnamefont{and} \bibinfo{author}{\bibnamefont{U.A.Wiedemann}},
  \bibinfo{journal}{Phys. \ Rev. \ Lett.} \textbf{\bibinfo{volume}{98}},
  \bibinfo{pages}{182301} (\bibinfo{year}{2007}).

\bibitem[{\citenamefont{Bodwin et~al.}(1995)\citenamefont{Bodwin, Braaten, and
  Lepage}}]{Bodwin:1994jh}
\bibinfo{author}{\bibfnamefont{G.~T.} \bibnamefont{Bodwin}},
  \bibinfo{author}{\bibfnamefont{E.}~\bibnamefont{Braaten}}, \bibnamefont{and}
  \bibinfo{author}{\bibfnamefont{G.~P.} \bibnamefont{Lepage}},
  \bibinfo{journal}{Phys. Rev.} \textbf{\bibinfo{volume}{D51}},
  \bibinfo{pages}{1125} (\bibinfo{year}{1995}), \eprint{hep-ph/9407339}.

\bibitem[{\citenamefont{Brambilla et~al.}(2004)}]{QWG_YellowReport}
\bibinfo{author}{\bibfnamefont{N.}~\bibnamefont{Brambilla}}
  \bibnamefont{et~al.} (\bibinfo{year}{2004}), \eprint{hep-ph/0412158}.

\bibitem[{\citenamefont{Ackermann et~al.}(2003)}]{Ackermann:2002ad}
\bibinfo{author}{\bibfnamefont{K.~H.} \bibnamefont{Ackermann}}
  \bibnamefont{et~al.}, \bibinfo{journal}{Nucl. Instrum Meth.}
  \textbf{\bibinfo{volume}{A499}}, \bibinfo{pages}{624} (\bibinfo{year}{2003}).

\bibitem[{\citenamefont{Beddo et~al.}(2003)}]{STAR_BEMC}
\bibinfo{author}{\bibfnamefont{M.}~\bibnamefont{Beddo}} \bibnamefont{et~al.},
  \bibinfo{journal}{Nucl. Instrum. Meth.} \textbf{\bibinfo{volume}{A499}},
  \bibinfo{pages}{725} (\bibinfo{year}{2003}).

\bibitem[{\citenamefont{Abelev et~al.}(2008{\natexlab{a}})}]{STAR_embedding}
\bibinfo{author}{\bibfnamefont{B.~I.} \bibnamefont{Abelev}}
  \bibnamefont{et~al.} (\bibinfo{year}{2008}{\natexlab{a}}),
  \eprint{arXiv:0808.2041}.

\bibitem[{\citenamefont{Abelev et~al.}(2008{\natexlab{b}})}]{bedangaPhi}
\bibinfo{author}{\bibfnamefont{B.~I.} \bibnamefont{Abelev}}
  \bibnamefont{et~al.} (\bibinfo{year}{2008}{\natexlab{b}}),
  \eprint{arXiv:0810.4979}.

\bibitem[{\citenamefont{Anderson et~al.}(2003)}]{STAR_TPC}
\bibinfo{author}{\bibfnamefont{M.}~\bibnamefont{Anderson}}
  \bibnamefont{et~al.}, \bibinfo{journal}{Nucl. Instrum. Meth.}
  \textbf{\bibinfo{volume}{A499}}, \bibinfo{pages}{659} (\bibinfo{year}{2003}).

\bibitem[{\citenamefont{Xu et~al.}(2008)}]{TPCReCalib}
\bibinfo{author}{\bibfnamefont{Y.-C.} \bibnamefont{Xu}} \bibnamefont{et~al.}
  (\bibinfo{year}{2008}), \eprint{arXiv:0807.4303}.

\bibitem[{\citenamefont{Adams et~al.}(2005{\natexlab{b}})}]{starTOFelectron}
\bibinfo{author}{\bibfnamefont{J.}~\bibnamefont{Adams}} \bibnamefont{et~al.},
  \bibinfo{journal}{Phys. Rev. Lett.} \textbf{\bibinfo{volume}{94}},
  \bibinfo{pages}{062301} (\bibinfo{year}{2005}{\natexlab{b}}).

\bibitem[{\citenamefont{Adare et~al.}(2007{\natexlab{c}})}]{Adare:2006kf}
\bibinfo{author}{\bibfnamefont{A.}~\bibnamefont{Adare}} \bibnamefont{et~al.},
  \bibinfo{journal}{Phys. Rev. Lett.} \textbf{\bibinfo{volume}{98}},
  \bibinfo{pages}{232002} (\bibinfo{year}{2007}{\natexlab{c}}).

\bibitem[{\citenamefont{Abe et~al.}(1997)}]{JpsiSpectra_CDF}
\bibinfo{author}{\bibfnamefont{F.}~\bibnamefont{Abe}} \bibnamefont{et~al.},
  \bibinfo{journal}{Phys. Rev. Lett.} \textbf{\bibinfo{volume}{79}},
  \bibinfo{pages}{572} (\bibinfo{year}{1997}).

\bibitem[{\citenamefont{Acosta et~al.}(2005)}]{JpsiSpectra_CDFII}
\bibinfo{author}{\bibfnamefont{D.~E.} \bibnamefont{Acosta}}
  \bibnamefont{et~al.}, \bibinfo{journal}{Phys. Rev.}
  \textbf{\bibinfo{volume}{D71}}, \bibinfo{pages}{032001}
  (\bibinfo{year}{2005}).

\bibitem[{\citenamefont{Sjostrand et~al.}(2006)\citenamefont{Sjostrand, Mrenna,
  and Skands}}]{Sjostrand:2006za}
\bibinfo{author}{\bibfnamefont{T.}~\bibnamefont{Sjostrand}},
  \bibinfo{author}{\bibfnamefont{S.}~\bibnamefont{Mrenna}}, \bibnamefont{and}
  \bibinfo{author}{\bibfnamefont{P.}~\bibnamefont{Skands}},
  \bibinfo{journal}{JHEP} \textbf{\bibinfo{volume}{05}}, \bibinfo{pages}{026}
  (\bibinfo{year}{2006}).

\bibitem[{\citenamefont{Tang}(2009)}]{zeboThesis}
\bibinfo{author}{\bibfnamefont{Z.}~\bibnamefont{Tang}}, Ph.D. thesis,
  \bibinfo{school}{University of Science and Technology and China}
  (\bibinfo{year}{2009}).

\bibitem[{\citenamefont{Adams et~al.}(2003)}]{ppUncertainty}
\bibinfo{author}{\bibfnamefont{J.}~\bibnamefont{Adams}} \bibnamefont{et~al.},
  \bibinfo{journal}{Phys. Rev. Lett.} \textbf{\bibinfo{volume}{91}},
  \bibinfo{pages}{172302} (\bibinfo{year}{2003}).

\bibitem[{\citenamefont{Nayak et~al.}(2003)\citenamefont{Nayak, Liu, and
  Cooper}}]{Nayak:2003jp}
\bibinfo{author}{\bibfnamefont{G.~C.} \bibnamefont{Nayak}},
  \bibinfo{author}{\bibfnamefont{M.~X.} \bibnamefont{Liu}}, \bibnamefont{and}
  \bibinfo{author}{\bibfnamefont{F.}~\bibnamefont{Cooper}},
  \bibinfo{journal}{Phys. Rev.} \textbf{\bibinfo{volume}{D68}},
  \bibinfo{pages}{034003} (\bibinfo{year}{2003}), \bibinfo{note}{and private
  communication}.

\bibitem[{\citenamefont{Artoisenet et~al.}(2008)}]{Artoisenet:2008fc}
\bibinfo{author}{\bibfnamefont{P.}~\bibnamefont{Artoisenet}}
  \bibnamefont{et~al.}, \bibinfo{journal}{Phys. Rev. Lett.}
  \textbf{\bibinfo{volume}{101}}, \bibinfo{pages}{152001}
  (\bibinfo{year}{2008}), \bibinfo{note}{and J.P. Lansberg private
  communication}.

\bibitem[{\citenamefont{Banner et~al.}(1982)}]{pion_UA2}
\bibinfo{author}{\bibfnamefont{M.}~\bibnamefont{Banner}} \bibnamefont{et~al.},
  \bibinfo{journal}{Phys. Lett.} \textbf{\bibinfo{volume}{B115}},
  \bibinfo{pages}{59} (\bibinfo{year}{1982}).

\bibitem[{\citenamefont{Adams et~al.}(2006)}]{scalingpi}
\bibinfo{author}{\bibfnamefont{J.}~\bibnamefont{Adams}} \bibnamefont{et~al.},
  \bibinfo{journal}{Phys. Lett.} \textbf{\bibinfo{volume}{B637}},
  \bibinfo{pages}{161} (\bibinfo{year}{2006}).

\bibitem[{\citenamefont{Adams et~al.}(2005{\natexlab{c}})}]{Lijuan}
\bibinfo{author}{\bibfnamefont{J.}~\bibnamefont{Adams}} \bibnamefont{et~al.},
  \bibinfo{journal}{Phys. Lett.} \textbf{\bibinfo{volume}{B616}},
  \bibinfo{pages}{8} (\bibinfo{year}{2005}{\natexlab{c}}).

\bibitem[{\citenamefont{Alper et~al.}(1975)}]{pion_proton_ISR}
\bibinfo{author}{\bibfnamefont{B.}~\bibnamefont{Alper}} \bibnamefont{et~al.},
  \bibinfo{journal}{Nucl. Phys.} \textbf{\bibinfo{volume}{B100}},
  \bibinfo{pages}{237} (\bibinfo{year}{1975}).

\bibitem[{\citenamefont{Antreasyan et~al.}(1979)}]{proton_FNAL}
\bibinfo{author}{\bibfnamefont{D.}~\bibnamefont{Antreasyan}}
  \bibnamefont{et~al.}, \bibinfo{journal}{Phys. Rev.}
  \textbf{\bibinfo{volume}{D19}}, \bibinfo{pages}{764} (\bibinfo{year}{1979}).

\bibitem[{\citenamefont{Albajar et~al.}(1991)}]{UA1Simulation}
\bibinfo{author}{\bibfnamefont{C.}~\bibnamefont{Albajar}} \bibnamefont{et~al.},
  \bibinfo{journal}{Phys. Lett.} \textbf{\bibinfo{volume}{B256}},
  \bibinfo{pages}{112} (\bibinfo{year}{1991}).

\bibitem[{\citenamefont{Kourkoumelis et~al.}(1980)}]{JpsiSpectra_ISR}
\bibinfo{author}{\bibfnamefont{C.}~\bibnamefont{Kourkoumelis}}
  \bibnamefont{et~al.}, \bibinfo{journal}{Phys. Lett.}
  \textbf{\bibinfo{volume}{B91}}, \bibinfo{pages}{481} (\bibinfo{year}{1980}).

\bibitem[{\citenamefont{Clark et~al.}(1978)}]{xT_scaling_history1}
\bibinfo{author}{\bibfnamefont{A.~G.} \bibnamefont{Clark}}
  \bibnamefont{et~al.}, \bibinfo{journal}{Phys. Lett.}
  \textbf{\bibinfo{volume}{B74}}, \bibinfo{pages}{267} (\bibinfo{year}{1978}).

\bibitem[{\citenamefont{Angelis et~al.}(1978)}]{xT_scaling_history2}
\bibinfo{author}{\bibfnamefont{A.~L.~S.} \bibnamefont{Angelis}}
  \bibnamefont{et~al.}, \bibinfo{journal}{Phys. Lett.}
  \textbf{\bibinfo{volume}{B79}}, \bibinfo{pages}{505} (\bibinfo{year}{1978}).

\bibitem[{\citenamefont{Adler et~al.}(2004)}]{xT_scaling_history3}
\bibinfo{author}{\bibfnamefont{S.~S.} \bibnamefont{Adler}}
  \bibnamefont{et~al.}, \bibinfo{journal}{Phys. Rev.}
  \textbf{\bibinfo{volume}{C69}}, \bibinfo{pages}{034910}
  (\bibinfo{year}{2004}).

\bibitem[{\citenamefont{Bedjidian et~al.}(2004)}]{Bedjidian:2004gd}
\bibinfo{author}{\bibfnamefont{M.}~\bibnamefont{Bedjidian}}
  \bibnamefont{et~al.} (\bibinfo{year}{2004}), \bibinfo{note}{and R. Vogt
  private communication}.

\bibitem[{\citenamefont{Adler et~al.}(2002)}]{Adler:2002xw}
\bibinfo{author}{\bibfnamefont{C.}~\bibnamefont{Adler}} \bibnamefont{et~al.},
  \bibinfo{journal}{Phys. Rev. Lett.} \textbf{\bibinfo{volume}{89}},
  \bibinfo{pages}{202301} (\bibinfo{year}{2002}).

\bibitem[{\citenamefont{Adare et~al.}(2008)}]{PHENIX_CuCu}
\bibinfo{author}{\bibfnamefont{A.}~\bibnamefont{Adare}} \bibnamefont{et~al.},
  \bibinfo{journal}{Phys. Rev. Lett.} \textbf{\bibinfo{volume}{101}},
  \bibinfo{pages}{122301} (\bibinfo{year}{2008}).

\bibitem[{\citenamefont{Gunji}(2008)}]{hydro+hotWind}
\bibinfo{author}{\bibfnamefont{T.}~\bibnamefont{Gunji}}, \bibinfo{journal}{J.
  Phys.G: Nucl. Part. Phys.} \textbf{\bibinfo{volume}{35}},
  \bibinfo{pages}{104137} (\bibinfo{year}{2008}).

\bibitem[{\citenamefont{Wicks et~al.}(2007)}]{Wicks:2005gt}
\bibinfo{author}{\bibfnamefont{S.}~\bibnamefont{Wicks}} \bibnamefont{et~al.},
  \bibinfo{journal}{Nucl. Phys.} \textbf{\bibinfo{volume}{A784}},
  \bibinfo{pages}{426} (\bibinfo{year}{2007}), \bibinfo{note}{and W. A.
  Horowitz private communication}.

\bibitem[{\citenamefont{Adil and Vitev}(2007)}]{Adil:2006ra}
\bibinfo{author}{\bibfnamefont{A.}~\bibnamefont{Adil}} \bibnamefont{and}
  \bibinfo{author}{\bibfnamefont{I.}~\bibnamefont{Vitev}},
  \bibinfo{journal}{Phys. Lett.} \textbf{\bibinfo{volume}{B649}},
  \bibinfo{pages}{139} (\bibinfo{year}{2007}), \bibinfo{note}{and I. Vitev
  private communication}.

\bibitem[{\citenamefont{Xu}(2002)}]{xmXu}
\bibinfo{author}{\bibfnamefont{X.-M.} \bibnamefont{Xu}},
  \bibinfo{journal}{Nucl. Phys.} \textbf{\bibinfo{volume}{A697}},
  \bibinfo{pages}{825} (\bibinfo{year}{2002}).

\bibitem[{\citenamefont{Adams et~al.}(2005{\natexlab{d}})}]{STAR_diHadron}
\bibinfo{author}{\bibfnamefont{J.}~\bibnamefont{Adams}} \bibnamefont{et~al.},
  \bibinfo{journal}{Phys. Rev. Lett.} \textbf{\bibinfo{volume}{95}},
  \bibinfo{pages}{152301} (\bibinfo{year}{2005}{\natexlab{d}}).

\end{thebibliography}
\end{document}